\documentclass[letter]{jpconf}
\usepackage{graphicx}
\begin{document}
\title{A Signature of Colour-Superconducting Dark Matter?}

\author{Kyle Lawson}

\address{ Department of Physics and Astronomy, University of British 
Columbia, Vancouver, BC, V6T 1Z1, CANADA}

\ead{klawson@phas.ubc.ca}

\begin{abstract}
I describe a novel dark matter candidate in which the dark matter is 
composed of macroscopically large ``nuggets" of ordinary quarks 
and antiquarks in a colour-superconducting phase. The physical properties of these  
objects are described entirely by \textsc{qcd} and the principles of 
condensed matter physics. An understanding of these properties 
allows for predictions of their interactions with the ordinary 
visible matter of the galaxy and leads to several testable consequences 
of the model. The spectrum arising from these interactions is entirely fixed by 
quite general predictions about the structure of dense quark matter and 
allows for no tuning of parameters. In this talk I present 
the results of a detailed Thomas-Fermi calculation which demonstrates the 
plausibility that the annihilation of galactic matter incident on a dark matter 
nugget may be responsible for both the galactic 511 keV line and a broad 
MeV scale continuum present in the galactic spectrum measured by 
\textsc{comptel}.
\end{abstract}

\section{Introduction}
The dominant fraction of the mass of the galaxy is comprised of dark matter 
which dominates over the visible matter by a ratio of 5:1. To this point there 
is no accepted theory as to the microscopic composition of dark matter though a 
wide range of theories have been proposed. The standard model contains no 
particle with the correct properties to account for the 
dark matter. Therefore, if dark matter does represent a fundamental particle 
it will necessarily require the expansion of the field content of the standard model. Here I 
discuss an alternative possibility in which the dark matter is not a new 
fundamental field but is instead a composite object carrying a baryonic 
charge in the range $10^{20} - 10^{30}$. In this model baryogenesis occurs 
through a charge separation mechanism occurring at the \textsc{qcd} phase 
transition which compresses quarks and antiquarks into dense 
colour-superconducting (CS) nuggets . This nugget formation mechanism is 
slightly more efficient for antiquarks  than for quarks.  As 
there are more antiquarks locked up in the CS nuggets there will be a corresponding 
excess in free quarks which then hadronize making up the visible matter of the 
universe while the nuggets comprise the dark matter.  While governed 
entirely by \textsc{qcd} physics, the details of nugget formation  
are strongly dependent on the phase structure of high density \textsc{qcd} 
which is not well understood at present \cite{Zhitnitsky:2002qa}. Rather than attempting to directly 
calculate the details of nugget formation I instead take a phenomenological 
point of view. Under the assumption that the observer dark matter density 
is entirely in the the form of CS quark matter several 
non-trivial tests of the model may be made based on emissions which must 
necessarily accompany these objects. Many of the results of this work are based on
\cite{Lawson:2007kp} and \cite{FLZ} which should be consulted for further
details.

\section{Nugget Properties}
As mentioned above the properties of the dark matter nuggets will be 
determined entirely by \textsc{qcd} physics. However, the complexity 
of the calculations involved leave several properties highly uncertain.
Fortunately, numerical studies of the properties of high density quark matter 
allow us to make several statements about the interaction spectrum of these 
objects which are largely independent of the precise details. 

\subsection{Colour-Superconductivity}
At asymptotically large densities the ground state of \textsc{qcd} 
is not nuclear matter but a phase in which the quarks condense into 
Cooper pairs. Matter in this phase is referred to as a colour-superconductor
and is conceptually similar to conventional superconductivity. 
The formation of quark Cooper pairs favours equal 
numbers of up and down quarks and thus the quark 
matter will carry a net electric charge, positive 
in the case of quarks and negative in the case of antiquarks. Electrical 
neutrality is enforced by a the presence of leptons (either electrons or positrons.) 
While the quark matter surface is very sharp, its scale being set by strong force 
interactions and thus at the fermi scale, the leptons are only electromagnetically 
bound and are thus distributed over a larger region. In astrophysical contexts 
such a layer of leptons is referred to as an electrosphere. While the exact value 
depends on the phase of quark matter realized the lepton chemical potential 
near the quark surface is quite generally in the range from 
$\mu_0 \sim 10 - 100$ MeV. From the surface the lepton chemical 
potential will extrapolate to zero far from the nugget.
\subsection{Dark Matter}
The idea that dark matter, with its evidently small interaction rates, 
could be composed of macroscopic, strongly interacting objects 
may seem counterintuitive. In this model it is not the small interaction 
cross section that makes the nuggets ``dark" but the very small number 
densities required to explain the observed dark matter density. The 
scale of all observable properties will be proportional to the number 
density of the nuggets
which is small enough to avoid most observational constraints 
even if they are near the lowest possible mass (and therefore 
largest number density.) The very low number densities make the 
prospects for earth based detection with all but the largest detectors 
unlikely.
\section{Emission Mechanisms}
The primary observational signatures of this dark matter model will 
be photons produced by the annihilation of visible galactic matter 
with the components of the antiquark nuggets. As discussed above 
these emission mechanisms will be strongly suppressed by the small 
nugget number density. It is only in regions where both the visible 
and dark matter densities are large that a detectable signature may 
be found. As such, the galactic centre is the most likely source for 
emissions with a flux large enough to be detectable. 

In the following sections I will discuss the emission spectrum resulting 
from the annihilation of galactic electrons in the electrosphere of an 
antiquark nugget. This process has the benefit of being governed by 
simple \textsc{qed} physics from which a clear spectrum can be extracted. 
However, in order to describe the resulting annihilation spectrum it is necessary to 
first determine the properties of the electrosphere.
\subsection{The Electrosphere}
The density of leptons near the surface of the quark nugget will 
be determined by balancing the electrical attraction of the leptons 
to the surface with the degeneracy pressure which supports the 
electrosphere. The electric potential will satisfy the Poisson 
equation,
\begin{equation}
\label{eq:Poisson}
\nabla^2 \phi(r) = -4\pi e n(r).
\end{equation}
Where n is the lepton number density. The charge distribution 
must be supported by degeneracy pressure such that
$e\phi(r) = -\mu(r)$. The expressions \ref{eq:Poisson} 
combined with the Fermi distribution expression for 
lepton number density my be solved numerically in the 
Thomas Fermi approximation to obtain a mean field 
density distribution for the electrosphere. The full details of 
this calculation are beyond the scope of this work but are 
presented in \cite{FLZ}. The important 
conclusion of this calculation is that 
there will exist a relatively large region of the 
electrosphere (roughly equivalent to the radius of the 
nugget itself) in which the leptons have densities at the 
atomic scale and non-relativistic velocities. Very near 
the quark surface the leptons will have ultrarelativistic 
momenta and large densities related to the length 
scale $\mu_0^{-1}$. 
\subsection{Positronium Formation}
If the centre of mass momentum between the 
incident galactic electron and the electrosphere 
positron is sufficiently small the most likely annihilation 
channel is through an intermediate state positronium. 
This positronium atom then decays through either a 
pair of back to back 511 keV photons or a three photon continuum 
the details of which are well understood. At leading order the 
wavefunction overlap between an incident $e^+e^-$ pair 
with centre of mass momentum q and 
a positronium atom is $(1+\frac{q}{m\alpha})^{-4}$. So that the 
rate of positronium formation falls off rapidly with growing momentum,
note that the characteristic momentum scale for this fall off is 
$m\alpha$. The exact cross section for positronium formation 
does not have a simple analytic expression and requires a 
numerically complicated summation over all positronium 
excitation levels. For present purposes it will suffice to note that 
the typical length scale of a ground state positronium atom 
is the Bohr radius obtained by solving the Schrodinger equation 
for an $e^+e^-$ bound state, $a_{Ps} = \frac{2}{m\alpha}$. 
This length scale gives a reasonable estimation for the 
formation cross section. 
Combining this expression with the density distribution of 
the electrosphere allows for the calculation of the strength 
the 511 keV line produced by galactic electrons incident on 
a dark matter nugget. However, the scale of this spectrum 
depends on the rate of collisions between visible and dark 
matter averaged along a given line of sight. This number is 
sensitive to the total baryonic charge of the nuggets and the 
exact distribution of galactic dark matter both of which are 
highly uncertain. 
\subsection{Direct Annihilation}
As a galactic electron approaches the quark matter surface 
the positrons it encounters carry an ever larger momentum 
and positronium formation becomes disfavoured. Instead 
annihilation occurs through the direct $e^+e^- \rightarrow 
2\gamma$ channel described by a simple \textsc{qed} 
calculation. Again the annihilation cross-section and rate 
may be calculated in order to determine the resulting 
spectrum. The details of this spectrum were presented in 
\cite{Lawson:2007kp}, for this work the only necessary 
details are that the spectrum covers a large range from 
below the electron mass up to the chemical potential at the 
level where annihilation occurs. It should be noted that once 
the chemical potential becomes larger than$\sim 20MeV$ the 
annihilation rate becomes large enough that virtually no 
electrons are able to penetrate any deeper. Thus, even if the 
surface chemical potential is much larger (even above the largest 
predicted values of $\sim 100MeV$) the annihilation spectrum 
will still have a maximum energy well below this level. 
The only unknown parameter appearing in the resulting spectrum 
is the total line of sight rate of collisions between visible and dark matter 
which also appeared in the positronium spectrum. While the 
present data does not allow for a prediction of the magnitude 
of the spectrum it does allow the prediction of the relative 
strengths of the 511 keV line and the direct annihilation 
continuum. 
\section{Observations}
The \textsc{integral} satellite has measured the flux of 511 keV
photons from the galactic centre \cite{Jean:2003ci}.
This flux is found to be strongly correlated with the galactic 
bulge with possible evidence for a significantly weaker disc 
component. Along the line of sight of the galactic centre the 
observed flux is measured to be $\frac{d\Phi}{d\Omega}\simeq 0.025 $ 
photons cm$^{-2}$s$^{-1}$sr$^{-1}$  \cite{Jean:2003ci}. 
Known astrophysical processes seem unable to explain the 
observed strength.  
The situation with emission in the MeV range is more complicated 
than that of the 511 keV line. This energy range has been observed 
by the \textsc{comptel} satellite but with a lower energy resolution .  
While some background processes in this range 
are easily understood others, particularly cosmic rays scattering off 
the interstellar medium have much larger uncertainties.  When the 
spectrum predicted from cosmic ray and astrophysical processes
is subtracted from the \textsc{comptel} data there seems to remain 
an unexplained diffuse component of the spectrum associated with 
the galactic centre \cite{Strong}. 
\section{Results}
If both the 511 keV line and the diffuse MeV continuum have as their 
origin the annihilation of galactic electrons within the electrosphere 
of an antiquark nugget then several correlations must exist between
these seemingly unrelated spectral features. The two sources must 
be spacially correlated though current data does not allow for any strong 
statements on such a relation. More interestingly the 
total magnitude of the two emission sources must be related by the 
branching fraction between positronium formation and direct annihilation. 
This fraction may be evaluated by averaging the annihilation rates 
for each process over the electron's path through the electrosphere. 
Numerically this gives an MeV continuum with roughly one tenth the 
total intensity as the positronium decay spectrum. When we use this 
value to set the scale of the MeV spectrum the result provides a remarkably 
good fit to the \textsc{comptel} data. 
\section{Conclusions}
If the dark matter is composed of nuggets of colour-superconducting 
quark matter than many of its properties can be determined based on 
well understood physical principles. In particular the structure of the 
surrounding electrosphere may be determined based on calculations 
in \textsc{qed} and condensed matter physics. Once this structure is 
established the emission spectrum resulting from the annihilation of 
galactic electrons immediately follows. Both the spectral density and 
the branching fraction between positronium formation and direct annihilation 
are entirely fixed with no tunable parameters. Only the overall scale of
the emission remains undetermined due to large uncertainties in both current 
data and theory related to high density \textsc{qcd}. The assumption that  
the observed galactic 511 keV line is associated with positronium formation 
and subsequent decay allows us to fix the net annihilation rate and the 
net strength of MeV scale continuum. It is found that the resulting spectrum 
falls precisely in the 1-30MeV range where the \textsc{comptel} data seems 
to require a new emission source. If the predicted MeV emission scale had been 
found to be any larger or spread over a wider energy range this model 
would have been immediately ruled out. Instead, the \textsc{comptel} data 
would seem to require a new emission source with precisely the amplitude 
and energy distribution generated from electron annihilation 
in the quark nugget dark matter scenario. Higher resolution data and a better 
understanding of astrophysical backgrounds will allow for further testing of 
this model. In particular both the 511 keV line and the MeV continuum 
should be found to have the same spacial distribution corresponding to 
the product of the visible and dark matter densities.

\end{document}